# Unraveling effects of competing interactions and frustration in vdW ferromagnetic $Fe_3GeTe_2$ nanoflake devices

Rajeswari Roy Chowdhury,[1][*] Daiichi Kurebayashi,[2] Jana Lustikova,[3] Oleg A. Tretiakov,[2] Shunsuke Fukami,[3,4,5,6,7][†] Ravi Prakash Singh,[1][†] Samik DuttaGupta,[3,4,8][†]

[1]*Department of Physics, Indian Institute of Science Education and Research Bhopal, Bhopal, Madhya Pradesh 462066. India*
[2]*University of New South Wales, Sydney, Australia*
[3]*Center for Science and Innovation in Spintronics, Tohoku University, 2-1-1 Katahira, Aoba-ku, Sendai 980-8577, Japan*
[4]*Laboratory for Nanoelectronics and Spintronics, Research Institute of Electrical Communication, Tohoku University, 2-1-1 Katahira, Aoba-ku, Sendai 980-8577, Japan*
[5]*Center for Innovative Integrated Electronic Systems, Tohoku University, 468-1 Aramaki Aza Aoba, Aoba-ku, Sendai 980-0845, Japan*
[6]*WPI-Advanced Institute for Materials Research, Tohoku University, 2-1-1 Katahira, Aoba-ku, Sendai 980-8577, Japan*
[7]*Inamori Research Institute for Science, Shijo, Shimogyo-ku, Kyoto 600-8411, Japan*
[8]*Saha Institute of Nuclear Physics, 1/AF Bidhannagar. Kolkata 700064, India*

[*] Present affiliation: Department of Physics, University of South Florida, Tampa, USA
Authors to whom correspondence should be addressed: *rajeswari1@usf.edu*, *rpsingh@iiserb.ac.in*, *duttagupta.samik@saha.ac.in*

**ABSTRACT**
**Two-dimensional (2D) van der Waals (vdW) magnets and devices have garnered significant attention owing to the stabilization of long range magnetic order down to atomic limit, and the prospect for novel quantum devices with unique functionalities. To achieve this objective, clarification of magnetotransport properties and understanding of the relevant interactions with lowering of dimensions are of extreme importance. Here, the magnetotransport properties of few atomic layer $Fe_3GeTe_2$ and $(Co_{0.25}Fe_{0.75})_3GeTe_2$ nanoflake devices have been investigated. Magnetotransport investigations with applied magnetic field along the easy-axis shows anomalous Hall effect, while that for applied magnetic field along the hard-axis reveals an unusual behaviour. Atomistic calculations considering the presence of antiferromagnetic, ferromagnetic and local symmetry-breaking interactions reveal critical role of magnetic frustration effect assisted by thermal fluctuations, leading to a non-zero scalar spin chirality manifesting in an unconventional Hall effect. The present result clarifies the underlying interactions in few-layer 2D vdW ferromagnetic material system, important for the understanding of non-collinear spin configurations in vdW magnets for 2D spintronic devices.**

Two-dimensional (2D) van der Waals (vdW) materials have drawn significant attention owing to their prospect for 2D magnetism, spintronic and quantum device applications [1-3]. The stabilization of a long-range magnetic order, seamless hetero assembly with various structures and the relative twisting of individual layers provide an exciting opportunity for exploration of novel physical properties and realization of quantum or spintronics devices down to the extreme monolayer limit [4-6]. Quasi two-dimensional metallic ferromagnet (FM) $Fe_3GeTe_2$ (FGT), an archetype 2D FM, has gathered significant attention owing to its relatively high Curie temperature (~ 220 K) [7,8], large anomalous Hall effect (AHE)[9], significant uniaxial magnetic anisotropy[10], controllable Curie temperature ($T_C$) [11], heavy fermion behavior [12,13], and unconventional magnetic ground state [14], down to the atomic limit[15]. Previous investigations in single-crystalline bulk have demonstrated the existence of complicated spin configurations including labyrinth domains, bubble domains as well as topologically protected skyrmion-like spin structures in this centrosymmetric system [16, 17]. Furthermore, previous experimental results have demonstrated an unconventional magnetic ground state in FGT comprising of an aggregate of skyrmion-like topological spin textures whose properties (*viz.* size, emergent magnetic field) are remarkably sensitive with doping either at the magnetic (Co-doped FGT) or non-magnetic site (As-doped FGT), qualitatively attributed to the emergence of symmetry-breaking interactions and/or magnetic frustration effect [18,19]. In this context, thickness dependent studies entails a critical aspect of their behavior, leading to the emergence of unconventional properties with reduction in layer number. With the lowering of dimensions, magnetotransport investigations under the application of out-of-plane magnetic fields in exfoliated flake devices supplemented by theoretical modelling and magnetic imaging down to few tens of nm thickness have revealed a multi-domain to single-domain type magnetization reversal phenomena, promising for the realization of 2D-based magnetic memory devices[20,21]. However, first principles calculation at reduced dimension predicts the existence of geometrically frustrated spin configuration, manifesting in an intraplanar antiferromagnet (AFM) state while the interplanar exchange interactions stabilize the overall FM ground state [22]. Intuitively, the presence of these competing interactions is expected to lead to non-zero scalar spin chirality (SSC), defined as the mixed product of three neighboring non-coplanar spins. For solid state systems, this SSC might result in an effective pseudo-magnetic field through the quantum mechanical Berry curvature in momentum space or can also serve as scattering sources for conduction electrons, phonons etc., leading to non-trivial electronic properties [23,24]. Consequently, for a material system like FGT, the competitive yet intricate balance between the various interactions along with thermal fluctuations might lead to fascinating effects at low dimensions compared to bulk. However, experimental and theoretical investigations focusing on interplay of competing interactions and magnetic frustration via longitudinal or transverse (*i.e.*, Hall-transport) magnetotransport under application of magnetic fields along easy and hard axes have remained elusive and demands careful investigation.

Here, we investigate temperature ($T$) dependent magnetotransport of uniaxial FM FGT and $(Co_{0.25}Fe_{0.75})_3GeTe_2$ ($Co_{0.25}$FGT) nanoflake devices. Our experimental results demonstrate an unusual cusp-like behavior in Hall resistivity, previously also observed in their bulk counterparts[18,19]. Complementary atomistic calculations reveal that this behavior can be attributed to arise from finite SSC, originating from stabilization of non-coplanar spin configurations, markedly different from that in the bulk. The obtained results demonstrate a novel magnetotransport behavior in vdW FM FGT, prospective for topological magnetism and clarification of the origin and realization of a variety of non-collinear spin textures at reduced dimensions.

Single crystals of FGT and $Co_{0.25}$FGT were grown by chemical vapor transport method with a temperature gradient of 750/650º C, reported in detail in our previous works [18,19]. Room temperature x-ray diffraction (XRD) patterns reveal the samples to be single phase, and the observed Bragg peaks can be indexed with (00*l*) peaks (see Figure S1(a), supplementary information). Compositional analyses were carried out using a scanning electron microscope (SEM) equipped with energy dispersive x-ray (EDX) spectroscopy (see Figure S1(b), supplementary information). FGT and $Co_{0.25}$FGT flakes were mechanically exfoliated in a glovebox in an inert atmosphere. We first exfoliated the flakes from bulk single crystals using a scotch tape method. Next, the exfoliated flakes were transferred to a viscoelastic polydimethylsiloxane (PDMS) stamp, and subsequently transferred to the Cr/Au prepatterns on $Si/SiO_2$ substrates using a dry transfer technique (see Figure S2, supplementary information). The entire process was monitored by an optical microscope enabling precise alignment of the flake on PDMS with the pre-patterns during contact. Cr/Au layers were fabricated using a lift-off photolithography technique. Prior to the transfer of the flakes on the pre-patterns, the substrates were cleaned using a low energy Ar-ion plasma. Magnetic properties of the bulk samples were investigated using a SQUID-VSM in the temperature ($T$) and magnetic field ($H$) range of 5-300 K and ±9 T, respectively. Transverse and longitudinal resistance measurements on the nanoflake devices under $H$ applied parallel or perpendicular to the surface of the exfoliated flakes were carried out using a Quantum Design physical property measurement (PPMS) system.

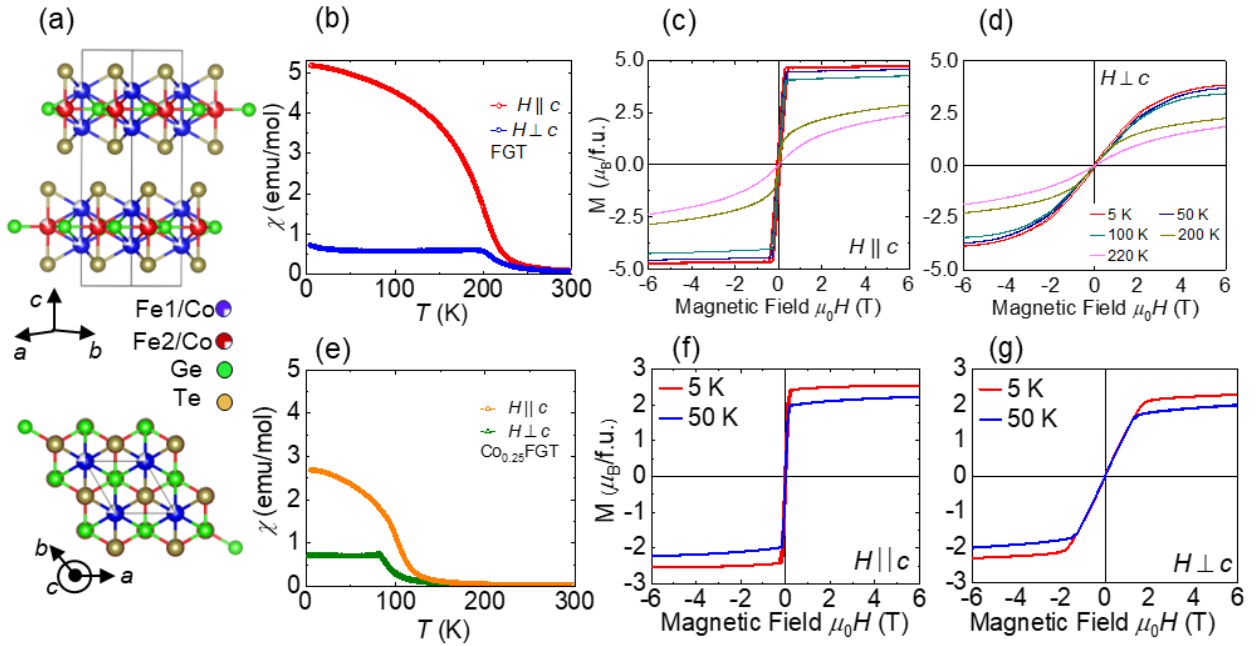

**FIG. 1.** (a) Top figure shows side view of the crystal structure of $Fe_3GeTe_2$ (FGT) and $(Co_{0.25}Fe_{0.75})_3GeTe_2$ ($Co_{0.25}$FGT). Bottom figure shows the crystal structure in the *ab*-pane with the *c*-axis being out-of-the plane direction. (b), (e) Magnetic susceptibility ($\chi_M$) versus temperature ($T$) under magnetic field $\mu_0 H = 0.5$T applied parallel to *c*-axis and perpendicular to *c*-axis for FGT and $Co_{0.25}$FGT, respectively, (c), (f) Applied magnetic field ($H$) dependence of magnetization ($M$) for FGT at various $T$ with $H \parallel c$-axis and $H \perp c$-axis, respectively. (f), (g) $M$ versus $H$ for $Co_{0.25}$FGT at indicated temperatures for $H \parallel c$-axis and $H \perp c$-axis, respectively.

Figure 1(a) shows the side (top row) and top (bottom row) view of the crystal structure of FGT and $Co_{0.25}$FGT,

crystallizing in a hexagonal structure with space group $P6_3/mmc$. Figures 1 (b) and (e) show the $T$ dependence of field-cooled (FC) magnetic susceptibility ($\chi_M$) for bulk single-crystalline FGT and Co$_{0.25}$FGT, under an applied magnetic field $\mu_0 H$ = 0.5 T, parallel (out-of-plane) or perpendicular (in-plane) to $c$-axis ($\mu_0$ is the permeability in vacuum), respectively. For FGT, from the derivative of $\chi_M$, we obtain an averaged Curie temperature ($T_C$) ≈ 201.63 (± 0.06) K, demonstrating the onset of ferromagnetic order. For Co$_{0.25}$FGT, $T_C$ reduces to ≈ 100.74 (± 0.05) K, consistent with our previous results showing monotonic decrease of $T_C$ with increased Co-doping [18]. The doping with Co also results in a suppressed bifurcation of $\chi_M$ between $H \parallel c$ and $H \perp c$-axes, attributed to the reduction of magnetic anisotropy. Figures 1(c), (d), (f), (g) show the magnetic hysteresis curves along the out-of-plane ($H \parallel c$) and in-plane ($H \perp c$) directions for Co$_{0.25}$FGT and FGT. The $M$-$H$ curves for $H \parallel c$-axis for both the samples tends to saturate at low magnetic field compared to $H \perp c$-axis, indicating magnetic easy-axis along the $c$ direction. To clarify the effect of doping on the magnetic anisotropy, $M$-$H$ curves for $H \perp c$-axis were utilized to determine first-order ($K_1$) and second-order ($K_2$) anisotropy constants by the Sucksmith-Thompson method. For Co$_{0.25}$FGT, we obtain $K_1(K_2)$ = 1.86 (0.26) × 10$^5$ J/m$^3$ at 5 K and $K_1(K_2)$ = 1.48 (0.08) × 10$^5$ J/m$^3$ at 50 K, indicating a significant weakening of magnetic anisotropy with Co-doping as compared to FGT[18,19], while retaining its uniaxial character (due to dominant $K_1$ contribution). The obtained results from the bulk FGT and Co$_{0.25}$FGT are later used for atomistic calculations to clarify the magnetotransport properties of the nanoflake devices.

Figure 2(a) shows the optical micrograph of a typical exfoliated device used for magnetotransport measurements. Figure 2(b) shows the atomic force microscopy image of the exfoliated flake devices, where the orange line indicates position of the line scan for flake thickness determination. From the height profile, we obtain a thickness of ~32.6 nm for Co$_{0.25}$FGT, and ~ 28.3 nm for FGT, respectively (Figure 2(c)). To clarify for any variation of electronic properties compared to their bulk counterparts, we measure the longitudinal resistance ($R_{XX}$) versus $T$ for Co$_{0.25}$FGT and FGT (Figure 2(d)) in a four-probe geometry. Both Co$_{0.25}$FGT and FGT nanoflakes exhibit metallic behavior along with a hump-like feature (indicated by arrows) around $T_C$, in good agreement with that determined from $\chi_M$ versus $T$. Compared to FGT, we also observe a reduction of $R_{XX}$ at similar device dimensions, similar to the results obtained for As-FGT and indicating an enhancement of metallicity with magnetic or non-magnetic doping [19,25].

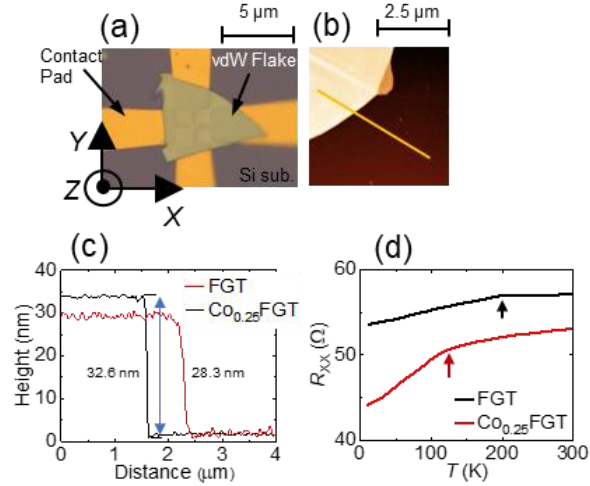

**FIG. 2:** (a) Optical micrograph of FGT or Co$_{0.25}$FGT device showing the pre-patterned contact pads with the exfoliated flake on top. Scale bar is 5 μm. (b) Atomic force microscope (AFM) image of the nanoflake at the edge. Black region corresponds to Si/SiO$_2$ substrate while the light-yellow region corresponds to the exfoliated flake. The orange line shows the position of the line scan for determination of flake thickness. Scale bar is 2.5 μm. (c) Height versus distance obtained from AFM image analysis for FGT and Co$_{0.25}$FGT. (d) Experimental results of $T$ dependence of longitudinal resistance ($R_{XX}$) for FGT and Co$_{0.25}$FGT. The black arrows correspond to the magnetic ordering temperature ($T_C$) in the nanoflake devices.

In order to have a deeper understanding on the underlying spin interactions and its impact under reduced dimensions in vdW FGT and Co-doped counterpart, we investigate the magnetotransport properties of Co$_{0.25}$FGT and FGT nanoflake devices. Figures 3(a), (b) show the experimental results of transverse resistance ($R_{XY}$) under applied $H_Z$ ($\parallel c$-axis) and current $I$ ($\parallel ab$-plane), *i.e.*, magnetic field applied along the magnetic easy-axes. For this configuration, we observe a sizable Hall resistance, previously attributed to anomalous Hall effect (AHE), possibly originating from topological nodal lines in the band structure with the magnetization pointing along the easy axis[26]. For FGT nanoflake devices, we observe a virtually square-shaped hysteresis with coercive field ($H_C$) ~ 94.7 mT at $T$ = 50 K, reminiscent of the hard magnetic behavior originating in the single domain regime[21]. For $Co_{0.25}FGT$, we also observe a similar square-shaped hysteresis curve with a much-reduced $H_C \sim 35$ mT at $T = 50$ K. These differences between FGT and $Co_{0.25}FGT$ can be attributed to the weakening of exchange (see Figure S3 of supplementary information), and anisotropy interactions associated with Co-doping, consistent with our previous results[18]. Interestingly, for applied field along the $x$ direction ($H_X \parallel I \perp c$-axis), we observe a significantly different behavior associated with an initial increment in the $R_{XY}$ magnitude followed by the emergence of a prominent cusp-like feature, and subsequent reduction of $R_{XY}$ magnitude at higher applied fields (Figures 3(c) and (d)). Note that while these are qualitatively similar to previous results on bulk FGT, Co and As-doped counterparts[18,19], as shown below, the dimensional reduction results in a completely different origin for the observed behavior which demands critical investigation.

Owing to the distinct domain wall pattern and magnetization dynamics in thin exfoliated flakes compared to that of bulk, we first calculate the total energy of the FGT and $Co_{0.25}FGT$ nanoflakes under varying thickness. We consider three types of domain configurations possible for a uniaxial magnet viz., single domain with moments aligned along the easy-axis, stripe domain configuration with alternate up/down directions of moments, and multidomain configuration with flux closure type of spin orientation. Using the experimentally determined values of micromagnetic parameters (exchange, anisotropy constant and saturation magnetization), we find that the single domain configuration corresponds to the lowest energy configuration for flake thickness lower than $\sim 60$ nm and $\sim 80$ nm, for FGT and $Co_{0.25}FGT$, respectively (see Figure S4 of supplementary information). Thus, for our nanoflake devices, this indicates the possibility of an exchange-dominated-type magnetization reversal governed by the Stoner-Wohlfarth model, for $R_{XY}$ versus $H_Z$ ($\parallel c$-axis) [27]. In contrast, the cusp-like feature can neither be explained by Stoner-Wohlfarth model or due to planar Hall effect under the applied $H_X$. Previous theoretical models indicate that a unit cell of FGT (and $Co_{0.25}FGT$) comprises of two monolayers of $Fe_3Ge$ sandwiched between two Te layers (Fig. 1(a)). The interactions between the intraplanar Fe atoms are of AFM type, while the FM ground state is realized by interplanar magnetic interactions among the magnetic atoms[28]. When viewed from the $c$-axis, the Fe atoms in a single monolayer form a triangular lattice resulting in finite SSC and serving as scattering sources for conduction electrons. Furthermore, intrinsic effects, such as presence of Fe deficiencies among the layers and/or interfacial effects might promote Dzyaloshinskii-Moriya interaction (DMI), previously shown to result in Néel-type chiral spin spirals or topological spin textures in centrosymmetric FGT[29], as has been observed in other vdW material $Fe_3GaTe_2$[30]. This combination of AFM and FM exchange, DMI and triangular lattice formed by Fe ions possibly renders magnetic fluctuations through magnetic frustration and/or stabilization of topological spin textures, leading to non-zero SSC and contributing to anomalies in magnetotransport in reduced dimensions. To further understand the possible origin behind $R_{XY}$ versus $H_X$, we utilize an atomistic spin model with the Heisenberg Hamiltonian (Eqn. 1) [31]:

$$H = -\sum_{i,j} J_{ij}\, S_i \cdot S_j + K \sum_i (S_i^z)^2 - \sum_i h_i \cdot S_i, \qquad (1)$$

where $J_{ij}$ is the exchange constant between $i^{th}$ and $j^{th}$ sites, $K$ is the uniaxial magnetic anisotropy constant (obtained from magnetization measurements), and $h_i$ is the applied magnetic field (see S5 of supplementary information). Interestingly, our calculations reveal that DM vector in the adjacent monolayers in a unit cell are in opposite directions, ruling out any contribution of DMI towards stabilization of topological spin textures in the exfoliated flake devices (see S6 of supplementary information). To capture any possible effect of magnetic frustration and/or fluctuation-induced non-coplanar spin configurations on the magnetotransport behavior, we calculate SSC (Eqn. 2) as

$$\chi = \frac{1}{2} \left| \sum_{i,j,k} b_{ijk} (S_i \cdot S_j \times S_k) \right|, \qquad (2)$$

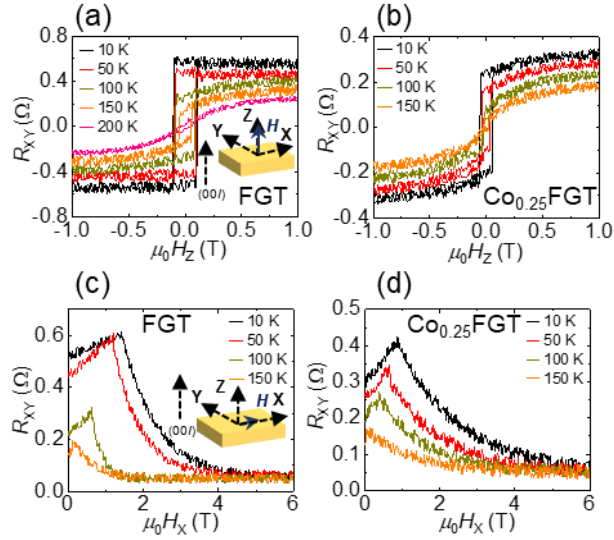

**FIG. 3:** (a) Experimental results of transverse resistance (or Hall resistance) ($R_{XY}$) versus $H_Z$ ($\parallel c$-axis) for FGT nanoflake devices at various $T$. Inset shows measurement configuration. (b) $R_{XY}$ versus $H_Z$ ($\parallel c$-axis) for $Co_{0.25}FGT$ nanoflake devices. Inset shows measurement configuration. (c) Experimental results of Hall resistance ($R_{XY}$) versus $H_X$ ($\perp c$-axis) for FGT nanoflake devices at various $T$. (d) $R_{XY}$ versus $H_X$ ($\perp c$-axis) for $Co_{0.25}FGT$ nanoflake devices.

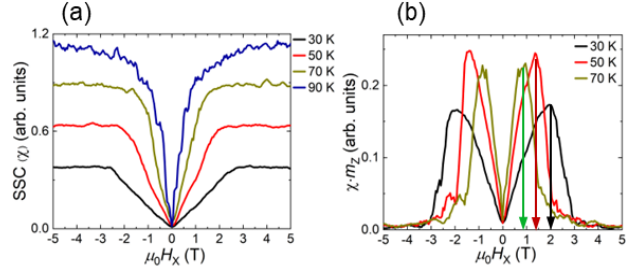

**FIG. 4:** (a) Scalar spin chirality (SSC) ($\chi$) versus in-plane magnetic field ($H_X$) ($\perp$ $c$-axis) at various $T$ for FGT. (b) SSC multiplied by out-of-plane component of magnetization ($\chi m_Z$) versus $H_X$ ($\perp$ $c$-axis) at various $T$ for FGT. Solid arrows correspond to the $H_X$ magnitude at which we observe an anomalous peak-like behavior in the experimental data.

where $b_{ijk}$ is the coefficient, $S_i, S_j, S_k$ are the normalized spin magnetic moments. Figure 4(a) shows $T$ dependence of SSC ($\chi$) versus applied in-plane magnetic field ($H_X$). Note that an increase in $H_X$ results in an enhancement of $\chi$, indicative of a significant effect of magnetic frustration. Furthermore, any emergence of non-trivial spin textures under $H_X$ is unlikely as it would correspond to additional features in $\chi$. An increase of $T$ also enhances $\chi$, hinting towards an enhanced magnetic frustration effect assisted by thermal fluctuations. The scalar product of $|\chi m_Z|$ would then correspond to the fictitious internal magnetic field contribution along the $c$-axis, responsible for the enhancement of $R_{XY}$ under $H_X$ under low magnetic fields. Further increase of $H_X$ results in the orientation of spins along the $ab$-plane, leading to decrease of SSC and fictitious magnetic field. To qualitatively compare the calculations with our experimental results, we calculate $|\chi m_Z|$ versus $H_X$ for different $T$ (Fig. 4(b)). Clearly, $|\chi m_Z|$ shows a peak where the Hall resistance ($R_{XY}$) shows a maximum under the application of $H_X$. Note that the peak value of $|\chi m_Z|$ appears at similar values of $H_X$ corresponding to the cusp-like behavior, in qualitative agreement with our experimental results. The threshold value of $H_X$ corresponding to the maximum of $|\chi m_Z|$ also shifts towards lower values with increasing $T$, justifying the predominance of magnetic frustration effect in FGT. Compared to FGT, the lower threshold values of $H_X$ for $Co_{0.25}FGT$ can be qualitatively attributed to arise from the weaker exchange and magnetic anisotropy interactions, resulting in a uniform magnetization state. Our study offers an in-depth understanding of the intricate role of magnetic frustration and/or fluctuation effects in few-layer vdW FMs, important for non-collinear spin configuration-based physics and future development of spintronic devices.

In summary, we have investigated the magnetotransport properties of FGT and $Co_{0.25}FGT$ nanoflake devices under out-of-plane ($H_Z \parallel c$-axis) and in-plane ($H_X \perp c$-axis) magnetic fields. For both FGT and $Co_{0.25}FGT$ nanoflake devices, transverse resistivity measurements under the application of $H_Z$ show virtually square-shaped AHE behavior, with significant coercivity compared to their bulk single crystals, expected for uniaxial magnets with strong magnetic anisotropy. On the other hand, for applied $H_X$ ($\perp$ $c$-axis), we observe an anomalous magnetotransport behavior manifesting in a cusp-like behavior. The threshold magnetic field corresponding to this cusp-like feature is controllable with Co-doping. Atomistic calculations reveal cancellation of DMI among adjacent monolayers, while competitive AFM, FM interactions result in thermally assisted frustrated spin configuration manifesting in a non-zero SSC, leading to the observed behavior. The thickness of the nanoflakes can affect the Hamiltonian of these vdW systems *viz.* magnetic anisotropy, demagnetization energy, and Zeeman energy, leading to changes in the internal magnetic domain structures in comparison to their bulk counterparts. The present study hereby signifies the critical role of magnetic frustration and/or fluctuation effects on unconventional magnetotransport phenomena in few-layer vdW FMs, offering a compelling avenue for understanding of non-collinear spin-configurations in 2D magnets for development of advanced spintronic devices.

**Author Contributions:**
R. Roy Chowdhury: Problem conceptualization, crystal growth and characterization, nanoflake exfoliation, magnetotransport experiment, writing, reviewing
Daichi Kurebayashi: Atomistic calculations
Jana Lustikova: Nanoflakes exfoliation
Oleg A. Tretiakov: Supervision-atomistic calculations
Shunsuke Fukami: Supervision & reviewing
Ravi Prakash Singh: Supervision & reviewing
Samik DuttaGupta: Patterning, nanoflakes exfoliation, magnetotransport experiment, writing, reviewing

**ACKNOWLEDGEMENTS**
R. Roy Chowdhury acknowledges Department of Science and Technology (DST), Government of India, for financial support (Grant no. DST/INSPIRE/04/2018/001755). R. P. Singh acknowledges Science and Engineering Research Board (SERB), Government of India, for Core Research Grant CRG/2023/000817. S. Fukami acknowledges JSPS Kakenhi for financial support (Grant Nos. 19H05622, 20K15155, 24H00039, and 24H02235). O. A. Tretiakov acknowledges the support from the Australian Research Council (Grant Nos. DP200101027 and DP240101062), the NCMAS grant, and RIEC Cooperative Research Projects, Tohoku University. S. DuttaGupta and O. A. Tretiakov acknowledges RIEC Cooperative Research Projects, Tohoku University for financial support.